\documentclass[letterpaper, showpacs, prl, lengthcheck,
superscriptaddress]{revtex4}

\pdfoutput=1

\usepackage{mathptmx}
\usepackage{amsmath, amsfonts}
\usepackage{graphicx}
\usepackage{hyperref}
\usepackage{subfigure}

\begin{document}

\title{Chaos in cylindrical stadium billiards via a generic nonlinear
  mechanism}
\author{Thomas Gilbert}
\email{thomas.gilbert@ulb.ac.be}
\affiliation{Center for Nonlinear Phenomena and Complex Systems,
  Universit\'e Libre  de Bruxelles, C.~P.~231, Campus Plaine, B-1050
  Brussels, Belgium}
\author{David P.~Sanders}
\email{dps@fciencias.unam.mx}
\affiliation{Departamento de F\'isica, Facultad de Ciencias, Universidad
Nacional Aut\'onoma de M\'exico,  Ciudad Universitaria,
04510 M\'exico D.F.,
 Mexico}

\begin{abstract}
We describe conditions under which higher-dimensional billiard models
in bounded, convex regions are fully chaotic, generalizing 
the Bunimovich stadium to dimensions above two. An example is a
three-dimensional stadium bounded by a cylinder and several planes; 
the combination of these elements may give rise to defocusing,
allowing large chaotic regions in phase space. By
studying  families of marginally-stable periodic orbits that populate
the residual part of phase space, we identify conditions under which
a nonlinear instability mechanism arises in their vicinity. For
particular geometries, this mechanism rather induces stable
nonlinear oscillations, 
including in the 
form of whispering-gallery modes. 
\end{abstract}


\maketitle

Billiard models, in which a point particle moves freely between elastic
collisions with a fixed boundary, are a fertile source of ideas in 
physics \cite{gaspard:1998book} and mathematics \cite{Sinai:1991p978} alike.
They provide a basis for some of the fundamental
concepts of statistical mechanics, and are at the same time open 
to mathematically rigorous analysis \cite{Szasz:2000book}. In particular,
they are some of the best-motivated models which exhibit strong
chaotic dynamics. 

There are two distinct categories of chaotic billiards. The first is the
class of \emph{dispersing} billiards, the prototypical example of which is the
hard-sphere gas: the dynamics of $N$ hard spheres in a
three-dimensional box with elastic collisions is
equivalent to a point particle in a $3N$-dimensional space moving
uniformly outside a collection of spherical cylinders, with specular 
reflections at the boundary \cite{Chernov:2006p683}.
The mechanism giving rise to chaos in such billiards is that of
dispersion, where nearby trajectories separate at each collision
with a convex surface; this leads to an overall exponential  divergence
and to a complete spectrum of Lyapunov exponents. The system is then
said to be fully chaotic or \emph{hyperbolic}.

The second category is made up of \emph{defocusing} billiards, the most
well-known example
of which is the Bunimovich stadium \cite{Bunimovich:1974p930}. 
Here, chaos is due to a mechanism different from dispersion, namely that of
defocusing: the boundary of the stadium
curves inwards with respect to the particle, so that nearby trajectories 
initially focus after colliding with this boundary; however,  the distance
to the next collision is typically longer than the distance to the focal
point, so that they eventually defocus even more. This again 
leads to an overall exponential expansion in phase space and hence complete
chaoticity.

Defocusing billiards have attracted much attention in the physics
community, particularly in connection with quantum chaos
\cite{Stockmann:1999p979}, acoustic experiments in 
chaotic cavities \cite{Draeger:1997p917}, optical microcavity laser
experiments \cite{Friedman:2001p900, Harayama:2003p901}, and quantum
conductance experiments \cite{Marcus:1992p977}, to name but a few. 

The extent to which the defocusing mechanism works in dimensions beyond
two has, however, long remained unclear \cite{BunManyDimStadium,
  Wojtkowski:1990p931}. In spite of some recent progress in addressing this
problem \cite{Bunimovich:1998p164, Papenbrock:2000p293,
  Bunimovich:2006p213}, 
the models studied are rather non-generic, consisting of
higher-dimensional planar surfaces with spherical caps
\cite{Bunimovich:1998p164}, or of a three-dimensional cube with
two mutually perpendicular half-cylindrical caps  \cite{Papenbrock:2000p293};
moreover, only the latter model is convex. Another example of three- and
four-dimensional convex billiards with flat and spherical components
thought to be chaotic was investigated in \cite{Bunimovich:1996p302}, but
it has so far not proved amenable to a systematic treatment.

In this Letter, we consider cylindrical stadium billiards, by which we
mean higher-dimensional convex billiards based on cylindrically-shaped
structures 
cut by planar elements. We describe the conditions under which they can
give rise to chaos through a conjunction of linear and nonlinear
instabilities, thus demonstrating that hyperbolicity in defocusing
billiards is easier to obtain than was previously believed. 

A generic dynamical feature of these billiards is that periodic orbits come
in continuous parametric families, which are associated to the flat
directions of the cylinders parallel to their axes. Each such 
direction thus gives rise to a pair of parabolic eigenvalues in the
eigenvalue spectrum of the periodic orbits. By studying the dynamics along the
directions of the corresponding eigenvectors at a nonlinear level, we
describe a mechanism by which families of periodic orbits can or cannot be
stabilized along these directions. Applying this method to
three-dimensional cylindrical stadium billiards, we establish the existence
of a large class of fully-chaotic convex billiards where 
this mechanism acts as a repulsive force. On the contrary, when it acts as a
restoring force, we identify stable modes of nonlinear oscillations, taking
the form of bouncing-ball orbits and whispering-gallery modes.

\begin{figure}[htb]
  \centering 
  \subfigure[]{
    \includegraphics[width=0.21\textwidth]{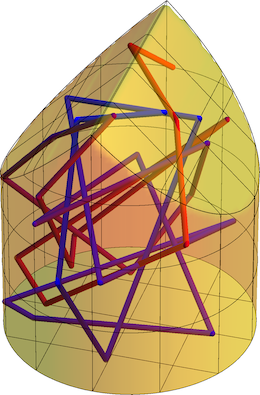}
    \label{fig.3dstadium2planes}
  }
  \hfill
  \subfigure[]{
    \includegraphics[width=0.21\textwidth]{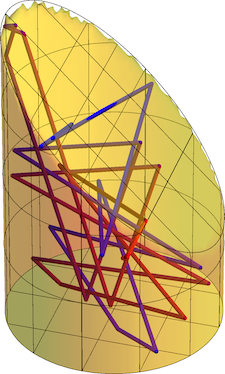}
    \label{fig.3dstadium1plane}
  }
  \caption{(Color online) Three-dimensional stadia consisting of a
    circular cylinder cut by: 
    \subref{fig.3dstadium2planes} three planes, 
    one perpendicular and two intersecting at right angles and cutting the
    cylinder at angle $\pi/4$; and
    \subref{fig.3dstadium1plane} two planes, one
    perpendicular to the cylinder axis and the other at angle $\pi/4$.
    In each case, a typical trajectory is depicted, with varying shades as
    time progresses.
  } 
  \label{fig.3dstadium}
\end{figure}

The construction of our class of models is simplicity itself: in dimension
three, they are formed by cutting a cylinder with one or more flat planes
to form a convex region,  with the billiard dynamics taking place
\emph{inside} this region. Two simple examples, consisting of a
three-dimensional circular cylinder cut by two or three planes, are shown
in Fig.~\ref{fig.3dstadium}. In both cases, one of the planes is
perpendicular to the cylinder axis, which just serves to confine the
motion, without introducing any new dynamical features. The other planes,
however, are angled away from perpendicular.  

\begin{figure}[htb]
  \centering 
  \subfigure[\ Planar periodic orbit]{
    \includegraphics[width=0.21\textwidth]{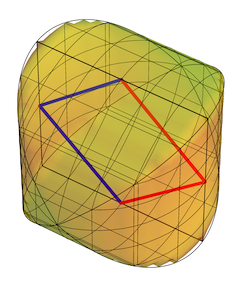}
    \label{fig.staboscPPO}
  }
  \hfill
  \subfigure[\ Helical periodic orbit]{
    \includegraphics[width=0.21\textwidth]{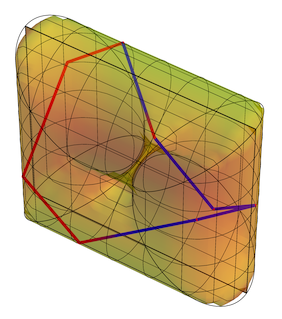}
    \label{fig.staboscHPO}
  }
  \caption{(Color online) Nonlinearly stable oscillations occur for specific
    geometries of the three-dimensional stadium shown in
    Fig.~\ref{fig.3dstadium1plane}. Two distinct types are 
    shown: \subref{fig.staboscPPO} Stable oscillations near the plane of
    symmetry of the billiard are found at negative height $h$ for a large class
    of periodic orbits; \subref{fig.staboscHPO} Approximations to helical
    periodic orbits whirl around the surface of the cylinder and can be
    stable  in small regions of phase space.
  }
  \label{fig.stabosc}
\end{figure}

One might  expect a cylindrical-shaped billiard to produce only
integrable behavior. This would indeed be so if the planes were all
perpendicular to the cylinder axis, leaving invariant the angular
momentum along the axis. However, when one of the planes is
oblique with respect to the cylinder axis, integrability is
lost and defocusing can take place. It turns out that this is already
sufficient to render the system of Fig.~\ref{fig.3dstadium2planes}
completely chaotic throughout its range of parameter values. There remains,
however, the possibility that stable oscillations take place in restricted
phase-space regions, whereby the system loses ergodicity. Billiards
like that of Fig.~\ref{fig.3dstadium1plane} may display such
behavior: depending on the height  $h$ of the oblique plane above the
cylinder base, nonlinearly stable 
oscillations are sometimes possible in confined regions of phase
space. Examples of such oscillations are shown in Fig.~\ref{fig.stabosc},
where we exploited the symmetry of the billiard to unfold it to a square cylindrical shape, 
by reflecting it multiple times in its planes, 

In general, we call a \emph{cylindrical stadium billiard} a bounded,
convex region made by cutting a cylinder with flat planes, such that at
least part of the boundary of the region is curved,
and such that the symmetries of the system are broken. This construction
easily extends to higher dimensions, as discussed below.

In cylindrical billiards, the problem of determining the existence of
stable elliptic islands is a fully nonlinear one. In the examples shown in
Fig.~\ref{fig.stabosc}, the phase space of the billiard map is
four-dimensional and every 
periodic orbit has four eigenvalues, two of which are parabolic ($\equiv
1$), corresponding to motion along the orbit's family. 
The remaining  pair can be either hyperbolic, in which case the
orbit is unstable, or elliptic, in which case it is marginally
stable. In order to assess the stability of the periodic orbit, it is then
necessary to analyze the motion along the parabolic eigenvectors which, at a
nonlinear level, is determined by the oscillations in the planes transverse
to the cylinder axes, associated with the pairs of elliptic
eigenvalues.  

The two parabolic eigenvectors correspond to the displacement along a
cylinder axis and its conjugate momentum. This displacement is identified
as parameterizing a continuous family of periodic orbits; we denote the
parameter by $\epsilon$. Apart from exceptional cases---for example, the
shape with negative cylinder height shown in
Fig.~\ref{fig.staboscPPO}---there is typically a critical value of the
parameter $\epsilon$ at which a bifurcation from elliptic to hyperbolic
regimes is found. This occurs  when the segments of the periodic orbit are
sufficiently long that defocusing takes place.  

\begin{figure}[tbh]
  \centering 
  \includegraphics[width=0.45\textwidth]{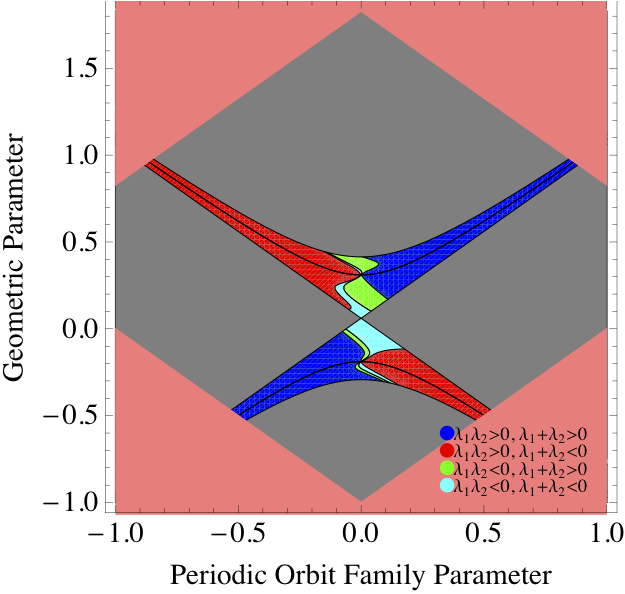}
  \caption{(Color) Analysis of the eigenvalues of the quadratic form
    (\ref{quadraticformw}) allows to identify the precise regions in
    parameter space where the periodic orbit shown in
    Fig.~\ref{fig.staboscHPO} is stable. The vertical axis 
    corresponds to the height $h$ of the cylindrical billiard
    \ref{fig.3dstadium1plane}, measured 
    at the lowest point of intersection of the oblique plane with the
    cylinder, and the horizontal axis to the parameter of the families of
    periodic orbits. Colors correspond to
    different regimes in parameter space: Light red indicates
    regions where the periodic orbit does not exist; gray 
    codes hyperbolic regions, where the orbit is unstable; and the remaining
    colors characterize the
    quadratic form $Q_w$ (\ref{quadraticformw}) in the region where the periodic
    orbits are marginally stable, according to the sign of its determinant
    (blue and red where positive; green and cyan where 
    negative) and trace (blue and green where positive; red and
    cyan where negative).  Similar results are found for 
    the orbit shown in Fig.~\ref{fig.staboscPPO}. 
    }
  \label{fig.nlstab}
\end{figure}

In the absence of nonlinear effects, oscillations around
a marginally-stable periodic orbit would always become unstable once the
perturbed orbit crossed the bifurcation point. There is, however, a
mechanism arising from the nonlinear corrections to the stability analysis,
which can be generically identified by the  properties of the eigenvalues
of a quadratic form that drives the oscillations of the 
momentum along the cylinder axis, i.e., the rate of displacement along the
family of periodic orbits. In the case of three-dimensional cylindrical
billiards, letting $w$ denote the momentum along the cylinder axis and
$\theta$ and $\xi$ the phase-space of the billiard map associated with the
motion in the transverse plane, the evolution  along a period of the
perturbation $\delta w$ of the momentum takes the form
\begin{equation}
  \delta w \mapsto \delta w + Q_w(\delta \theta, \delta\xi),
  \label{quadraticmapw}
\end{equation}
where $Q_w(\delta \theta, \delta\xi)$ is a quadratic form in the
variables $\delta \theta$ and $\delta \xi$, which we write as 
\begin{equation}
  Q_w(\delta \theta, \delta\xi) = 
  \Big(\delta \theta\quad\delta\xi\Big)
  \left(
    \begin{array}{cc}
      a_w & \frac{1}{2}b_w\\
      \frac{1}{2}b_w & c_w
    \end{array}
  \right)
  \left(
    \begin{array}{c}
      \delta\theta \\
      \delta\xi
    \end{array}
  \right),
  \label{quadraticformw}
\end{equation}
with coefficients $a_w$, $b_w$ and $c_w$ that are functions of the model's
geometric parameter $h$ and of the periodic orbit family parameter $\epsilon$.

Depending on the nature of the periodic orbit, the displacement
along the axis, which we interpret as a change in the parameter $\epsilon$,
may depend on $\theta$ and $\xi$ through  
a linear combination, or through a quadratic form similar to
(\ref{quadraticformw}). It is in any case a linear function of $w$, with a
coefficient which is independent of $\epsilon$. Given that
oscillations take place due to the motion in the $\theta$--$\xi$ plane,
nonlinear stability occurs whenever the perturbations along the
coordinate $w$, Eq.~(\ref{quadraticmapw}), act as a restoring force to
counteract these oscillations, so that they remain confined to the elliptic
range of the 
family. When these perturbations amplify the $\theta$--$\xi$ oscillations,
on the other hand, the nonlinear mechanism acts as a repulsive force and
destabilizes the family of periodic orbits.

These regimes can be analyzed in terms of the  properties of the
eigenvalues of the matrix in Eq.~(\ref{quadraticformw}). Having fixed the
geometry of the model, an interval of nonlinear stability in the family of
periodic orbits corresponds to values of the determinant and trace of the
quadratic form (\ref{quadraticformw}) such that $w$ takes the sign opposite
to that of the changes in the parameter $\epsilon$.
The results of this analysis are shown in Fig.~\ref{fig.nlstab} for 
the class of orbits shown in Fig.~\ref{fig.staboscHPO}. The regions which are 
found to be elliptic from the linear stability analysis are divided into
two symmetric tongues.  The nonlinear analysis, however,
shows that the upper tongue is unstable, and the lower one is stable only near
the middle of the range of the family, $\epsilon = 0$.

A detailed discussion may be found in \cite{GilbertSanders}. There, it
is shown that the periodic orbits under consideration are the crudest
approximation to a family of helical periodic orbits, whose periods
increase to infinity and which approximate an exact helix which can be
observed at height parameter $h = \pi/2-1$. The finite-approximation helical
periodic orbits all have small regions of stability, 
whose sizes in parameter space decrease in direct proportion to the inverse
of their periods. They can be interpreted as a sequence of
whispering-gallery modes of the billiard, indexed by their periods. The
nonlinear stability analysis of the planar orbits shown in
Fig.~\ref{fig.staboscPPO} yields very similar results. 

\begin{figure}[thb]
  \centering
  \includegraphics[width=0.45\textwidth,angle=0]{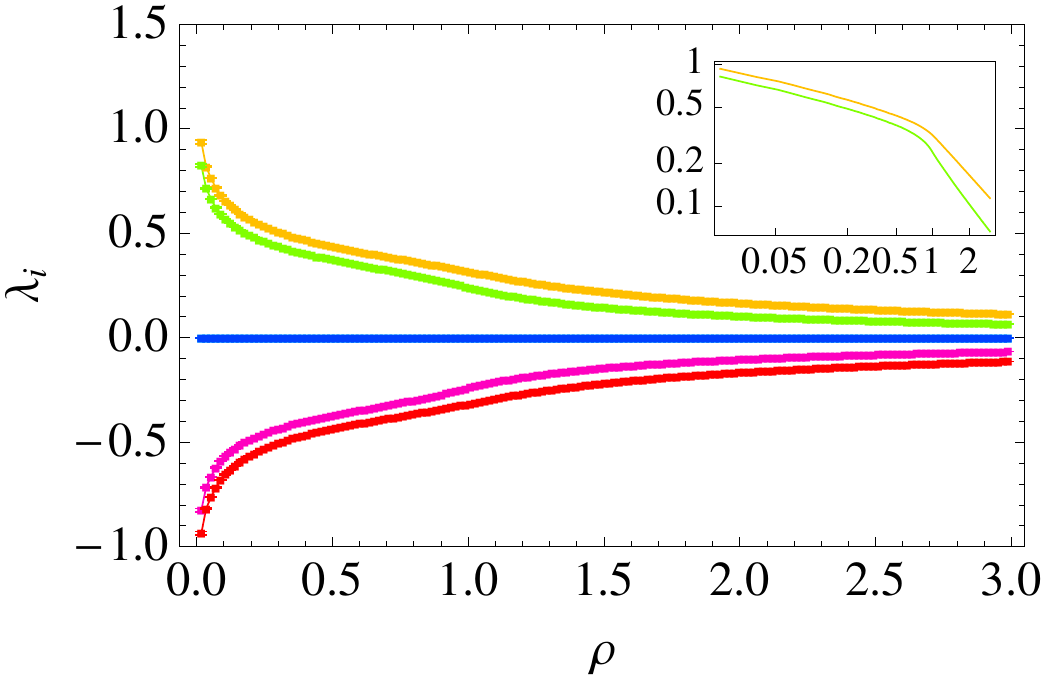}
  \caption{(Color online) Spectrum of Lyapunov exponents of the billiard shown
    in Fig.~\ref{fig.3dstadium2planes}. The parameter $\rho$ measures the
    ratio 
    between the radius of the cylinder and half its height. There are four
    non-zero exponents, arranged in two positive-negative pairs.
    The  inset shows a log-log plot of the positive exponents. As $\rho$
    increases, the exponents decrease with the frequency of collisions with
    the oblique planes.}
  \label{fig.3dstadiumlyap}
\end{figure}

These stable oscillations can easily be destroyed by changing the
geometry of the billiard. In particular, the billiard with two oblique
planes at right angle shown in Fig.~\ref{fig.3dstadium2planes} does not
admit nonlinearly stable oscillations like those reported above,
essentially because the perturbations will make collisions with both
oblique planes, causing them to become unstable. It is thus an example
of an ergodic, completely chaotic billiard in its whole parameter
range. That is, the Lyapunov exponents, which measure the separation 
rate of nearby 
trajectories, of almost every point in the $6$-dimensional phase space of
Cartesian coordinates are constant and nonzero, except for two
zero exponents associated with energy conservation and the
corresponding time translation. This claim is 
substantiated by the results of numerical computations of these exponents,
such as those shown in Fig.~\ref{fig.3dstadiumlyap}. 

Examples of the general class of billiards described above arise naturally
in models of interacting particles with flat surfaces, such as studied in
\cite{Gilbert:2008p354} in the context of heat conduction. 
For example, the model of Fig.~\ref{fig.3dstadium2planes} is equivalent to
a diatomic molecule with an interaction mediated by a massless string,  
and confined in an infinite two-dimensional channel. A systematic
exposition of this equivalence is given in Ref.~\cite{GilbertSanders}.

The generalization to higher-dimensional cylindrical billiards is
straightforward. For example, the same molecule confined to a square box
is equivalent to a cylinder in four dimensions with a two-dimensional
circular base and two perpendicular axes, cut by eight
hyperplanes. Similarly, a three-atom molecule such as studied in 
\cite{Papenbrock:2000p166} corresponds to a billiard in four
dimensions inside the  Cartesian product of two disks. In general,
higher-dimensional cylindrical billiards cut by hyperplanes are
expected to be chaotic. The existence of stable regions is rather
exceptional, being found only for systems admitting families of periodic
orbits with segments that remain close enough to the billiard
surface that defocusing does not take place, and where the generic nonlinear 
mechanism identified above comes into play as a restoring force, rather than
a repulsive one. 

A main conclusion of our work is that cylindrical stadium billiards
obtained by breaking the symmetry of cylindrical
structures by the insertion  of oblique planes easily yield fully
chaotic dynamics, in spite of the existence of marginally-stable regions in phase space. 
A nonlinear mechanism operating along the the flat directions of
the cylindrical surfaces may act as a repulsive force and prevent elliptic
periodic orbits 
from giving rise to stable oscillations in their vicinity.
In future work, we will explore the effect of this mechanism on cylindrical
billiards with different forms of bases; preliminary results suggest that even
cylindrical billiards with an elliptic base can be chaotic. 

An interesting perspective is that our billiards can be used as 
building blocks for spatially-extended periodic and non-periodic
structures, related to so-called track billiards
\cite{Bunimovich:2009p486}. These structures have straight segments and
are angled with respect to one  
another. Our results allow to identify geometries such that classical
dynamics within such structures is chaotic. This can be expected, for
instance, in a series  of straight tubes connected by joins. Such extended
structures display diffusive behavior, as will be reported in future
work.

The existence of such chaotic structures has further interesting
implications for physical systems and many potential applications, whether
in nanostructures, fluid dynamics, acoustic devices, or optical fibers,
where experiments would be 
possible. 

\begin{acknowledgments}
  The authors thank Carlangelo Liverani and Vered Rom-Kedar for helpful
  discussions. This research benefited from the joint support of FNRS
  (Belgium) and CONACYT (Mexico) through a bilateral collaboration
  project. The work of TG is financially supported by the Belgian Federal
  Government under the Inter-university Attraction Pole project NOSY
  P06/02. TG is financially supported by the Fonds de la Recherche
  Scientifique F.R.S.-FNRS.  DPS acknowledges financial support from
  DGAPA-UNAM grant  IN105209.
\end{acknowledgments}


\end{document}